\documentclass[12pt]{article}
\usepackage[dvips]{graphicx}
\def\thefootnote{\fnsymbol{footnote}}
\def\be{\begin{equation}}
\def\ee{\end{equation}}
\def\ba{\begin{eqnarray}}
\def\ea{\end{eqnarray}}

\begin{document}
\begin{titlepage}
\thispagestyle{empty}
\vskip0.5cm
\begin{flushright}
MS--TPI--98--15
% { \sc DRAFT}
\end{flushright}
\vskip0.8cm

\begin{center}
{\Large {\bf Zeroes in the Complex $\beta$ Plane}}
\end{center}
\begin{center}
{\Large {\bf Of 2D Ising Block Spin Boltzmannians}}
\end{center}
\vskip0.8cm
\begin{center}
{\large Klaus Pinn}\\
\vskip5mm
{Institut f\"ur Theoretische Physik I } \\
{Universit\"at M\"unster }\\ {Wilhelm--Klemm--Str.~9 }\\
{D--48149 M\"unster, Germany \\[5mm]
 e--mail: pinn@uni--muenster.de
 }
\end{center}
\vskip2.0cm
\begin{abstract}
\par\noindent
Effective Boltzmannians in the sense of the block spin renormalization
group are computed for the 2D Ising model. The blocking is done with
majority and Kadanoff rules for blocks of size 2 by 2. Transfer matrix
techniques allow the determination of the effective Boltzmannians as
polynomials in $u=\exp(4\beta)$ for lattices of up to 4 by 4 blocks.
The zeroes of these polynomials are computed for all non-equivalent
block spin configurations.  Their distribution in the complex $\beta$
plane reflects the regularity structure of the block spin
transformation.  In the case of the Kadanoff rule spurious zeroes
approach the positive real $\beta$ axis at large values of $\beta$.
They might be related to the renormalization group pathologies
discussed in the literature.
\end{abstract}
\end{titlepage}

\setcounter{footnote}{0}
\def\thefootnote{\arabic{footnote}}

\section{Introduction}

Regularity is at the heart of position space (block spin)
renormalization group. It is usually assumed, and of central
importance, that coupling constants of the block spin effective
Hamiltonian depend in a non-singular way on the parameters of the
original theory. There are, however, situations where this assumption
is not valid.  Pathological behaviour in renormalization groups of the
low temperature Ising model was first observed by Israel \cite{Israel}
and Griffith and Pearce \cite{Griffith1,Griffith3}.  An
extensive and rigorous analysis of regularity properties and
pathologies in Ising model block spin transformations was performed by
van Enter et al.\ \cite{vanEnter}.  The central observation is that in
certain situations the effective measure for the block spin theory
cannot be represented as $\exp(-H)$.  This means that the
effective measure is non-Gibbsian.  See also \cite{Bricmont} for a
careful analysis of the situation.
 
In this paper, I present some numerical results on the distribution 
of zeroes in the complex $\beta$ plane for block spin Boltzmannians
of the 2D Ising model. These Boltzmannians are partition functions 
with ``fixed'' block spins $\mu$, viz.\ 
\be
B(\mu) = \sum_{\sigma} P(\mu,\sigma) \, \exp[-\beta H(\sigma)] \, . 
\ee
$P(\mu,\sigma)$ encodes the blocking rule. 
Why should these zeroes provide interesting information? 
If the usual renormalization group assumptions are true, 
the zeroes of $B(\mu)$ should -- for all block spin configurations $\mu$ --
behave ``better'' than those of the full partition function. 
Note that the zeroes of the full partition function
approach the real axis 
at the critical point \cite{Itzykson}. This should {\em not}
happen for the zeroes of $B(\mu)$! Furthermore, one might
expect that the pathologies described in the literature are  
related to the distribution of zeroes close to the $\beta$ 
axis at large positive values.

This article is organized as follows: In section 2 the model 
notation is set up, and the blocking rules are defined. 
Section 3 gives a sketch of the transfer matrix technique
used to compute the polynomials. Section 4 summarizes the 
numerical results for the majority rule blocking. Observations 
for the Kadanoff blocking rule are reported in section 5. 
Conclusions follow.

\section{Model and Block Spin Definition} 

We deal with the 2D Ising model, with partition function 
\be
Z = \sum_{\sigma} \exp[- \beta H(\sigma)] \, , 
\ee 
where 
\be 
H(\sigma)= - \sum_{<i,j>} \sigma_i \sigma_j  \, . 
\ee 
The $\sigma_i$ assume values $\pm 1$ and 
are defined on a square lattice of extension $L$ by $L$, 
supplied with periodic boundary conditions. 
The energy  $H$ is a sum over all pairs of nearest neighbours.
In the infinite volume limit the model undergoes a second order 
phase transition at $\beta_c = \frac12 \ln(\sqrt{2}+1)= 0.4406868$. 
A block spin transformation with scale factor 2 is defined as 
follows. For $L$ even, the lattice is divided in blocks of 
size 2 by 2. Given the configuration of the $\sigma$-spins in 
a block $I$, a block spin $\mu_I$ is chosen with probability 
$P(\mu,\sigma)$. The majority rule is defined through 
\be 
P(\mu,\sigma) = \prod_{{\rm blocks} \; I} p_I(\mu_I,\sigma) \, , 
\ee 
with 
\be 
p_I(\mu_I,\sigma) = \left\{ 
\begin{array}{ll}
\frac12 & \mbox{\ if \ } \sum_{i\in I} \sigma_i = 0 \, ,   \\[2mm]
1       & \mbox{\ if \ } \mu_I \sum_{i\in I} \sigma_i > 0 \, ,    \\[2mm]
0       & \mbox{\ else } \, .  
\end{array}
\right.
\ee 
The so-called Kadanoff rule is 
\be
\label{kadanoff} 
p_I(\mu_I,\sigma) = 
\frac{\exp\left(\omega \mu_I \sum_{i\in I} \sigma_i\right)}
{2 \cosh\left(\omega \sum_{i\in I} \sigma_i\right)} \; . 
\ee 
In the limit $\omega \rightarrow \infty$ one recovers the majority rule. 

Given that $\sum_{\mu} P(\mu,\sigma) = 1$, the full partition function 
can be rewritten as 
\be
Z = \sum_{\mu} B(\mu) \, ,  
\ee
with 
\be
B(\mu) = \sum_{\sigma} P(\mu,\sigma) \, \exp[- \beta H(\sigma)] \, . 
\ee 
Usually, one aims at a parametrisation $B(\mu)= \exp[-H'(\mu)]$, 
where $H'$ is the effective Hamiltonian. 
Note, however, that this is impossible in the pathological situations
discussed in the literature \cite{vanEnter}.

\section{Computation of $B(\mu)$}

An exact computation of $B(\mu)$ as function of $\beta$
seems impossible. However, on lattices up to at least $L=8$
it can be determined by numerical transfer matrix
calculations. 
$L=8$ corresponds to 4 by 4 blocks, with 
$2^{16}=65,536$ block spin configurations $\mu$. 
We do not have to consider them all.
Configurations connected through 
global spin flip or geometric symmetries (reflections, shifts, rotations)
have the same effective Boltzmannian $B(\mu)$. A careful counting 
yields for the number of non-equivalent configurations $N(L)$
\begin{center}
\begin{tabular}{ccccc}
 $L$   & 4 &   6  &   8  &   10   \\ 
\hline 
$N(L)$ & 4 &  13 & 479 & 86056 \\
\end{tabular}
\end{center}
The result for $N(10)$ was taken from ref.\ \cite{Dudek}.

After multiplication with a constant 
pre-factor, $B(\mu)$ can be expressed
as a polynomial of order $L^2$ in $u=\exp(4\beta)$.
The coefficients of this polynomial can be computed by 
transfer matrix multiplication. 
In order to avoid notational complication, I give a sketch of
the method for the case of 3 by 3 blocks ($L=6$). 
The generalization to other values of $L$ is then obvious. 
Some of the notation is depicted in table~\ref{lam}. 
The effective Boltzmannian can be expressed as 
\be
\label{tramult}
B(\mu)= {\rm Tr}\left[
{\bf T} \cdot 
{\bf S}(\mu_7,\mu_8,\mu_9) \cdot
{\bf T} \cdot 
{\bf S}(\mu_4,\mu_5,\mu_6) \cdot
{\bf T} \cdot
{\bf S}(\mu_1,\mu_2,\mu_3) 
\right] \, . 
\ee
Here, 
{\bf S} is a $2^6$ by $2^6$ matrix, labeled by the Ising row 
configurations. 
It depends explicitly on the line configuration of 
prescribed block spins. E.g., 
the matrix elements of ${\bf S}(\mu_1,\mu_2,\mu_3)$ are 
\be
{\bf S}(\mu_1,\mu_2,\mu_3)_{\sigma,\tau}
= \exp\left[ 
\beta \sum_{i=1}^6 ( \sigma_i \sigma_{i+1} 
+ \tau_i \tau_{i+1} +  \sigma_i \tau_i) 
\right] \prod_{I=1}^3 p_I(\mu_I,\sigma,\tau) \, . 
\ee 
The matrix {\bf T}, which is also of size $2^6$ by $2^6$, is defined by 
\be 
{\bf T}_{\sigma,\tau }
= \exp\left[\beta \sum_{i=1}^6 \sigma_i \tau_i \right] \, .
\ee 
It is not difficult to represent 
the transfer matrix multiplications in eq.~(\ref{tramult}) in 
terms of operations on the coefficients of 
polynomials in $u$. The computer implementation of these 
operations form the basis for the results presented in this paper.  

\begin{table}
\small 
\begin{center}
\begin{tabular}{|lll|lll|lll|}
\hline
$\tau_.$ && $\tau_.$ & 
$\tau_.$ && $\tau_.$ & 
$\tau_.$ && $\tau_.$ \\[3mm]
 &\fbox{$\mu_7$}& & & \fbox{$\mu_8$} & & & \fbox{$\mu_9$} & \\[3mm]
$\sigma_.$ && $\sigma_.$ & 
$\sigma_.$ && $\sigma_.$ & 
$\sigma_.$ && $\sigma_.$ \\ 
\hline 
$\tau_.$ && $\tau_.$ & 
$\tau_.$ && $\tau_.$ & 
$\tau_.$ && $\tau_.$ \\[3mm] 
 &\fbox{$\mu_4$}& & & \fbox{$\mu_5$} & & & \fbox{$\mu_6$} & \\[3mm] 
$\sigma_.$ && $\sigma_.$ & 
$\sigma_.$ && $\sigma_.$ & 
$\sigma_.$ && $\sigma_.$ \\ 
\hline 
$\tau_1$ && $\tau_2$ & 
$\tau_3$ && $\tau_4$ & 
$\tau_5$ && $\tau_6$ \\[3mm] 
 &\fbox{$\mu_1$}& & & \fbox{$\mu_2$} & & & \fbox{$\mu_3$} & \\[3mm] 
$\sigma_1$ && $\sigma_2$ & 
$\sigma_3$ && $\sigma_4$ & 
$\sigma_5$ && $\sigma_6$ \\ 
 \hline
 \end{tabular}
\parbox[t]{.85\textwidth}
 {
 \caption[lam]
 {\label{lam}
\small
Notation for transfer matrix in the case $L=6$.
}}
\end{center}
\end{table}

\section{Results for the Majority Rule}

Let us start with some results for $L=4$. The four non-equivalent 
configurations, called $c\#1 \dots c\#4$, are specified
in the head of table~\ref{coeffl4}. In the columns we quote 
the coefficients $B_k(\mu)$ of the polynomial 
$B(\mu) = \sum_{k=0}^{L^2} B_k(\mu) \, u^k$.

\begin{table}
\newcommand{\cfA}{$\begin{array}{cc} + & - \\ - & + \end{array}$}
\newcommand{\cfB}{$\begin{array}{cc} + & + \\ - & - \end{array}$}
\newcommand{\cfC}{$\begin{array}{cc} + & + \\ + & - \end{array}$}
\newcommand{\cfD}{$\begin{array}{cc} + & + \\ + & + \end{array}$}
\small
\begin{center}
\begin{tabular}{|r|r|r|r|r|}
\hline
    & c\#1\phantom{xx} 
    & c\#2\phantom{xx}   
    & c\#3\phantom{xx}  
    & c\#4\phantom{xx}   \\
\hline 
$k$ & \cfA & \cfB   & \cfC  & \cfD   \\
\hline 
0  &     2 &      2 &     2 &      2 \\ 
1  &     0 &      0 &     0 &      0 \\ 
2  &    32 &     32 &    32 &     32 \\  
3  &    96 &     64 &    64 &     32 \\  
4  &   544 &    384 &   416 &    416 \\
5  &  2336 &   1728 &  1728 &   1120 \\  
6  &  9360 &   6336 &  6560 &   5232 \\ 
7  & 19712 &  12960 & 13344 &   9536 \\ 
8  & 23674 &  20906 & 20570 &  16426 \\  
9  &  8224 &  14592 & 14112 &  14688 \\  
10 &  1456 &   7360 &  6720 &  10448 \\
11 &    96 &   1120 &  1696 &   4704 \\   
12 &     4 &     52 &   260 &   2244 \\    
13 &     0 &      0 &    32 &    384 \\       
14 &     0 &      0 &     0 &    256 \\  
15 &     0 &      0 &     0 &      0 \\ 
16 &     0 &      0 &     0 &     16 \\                        
\hline 
 \end{tabular}
\parbox[t]{.85\textwidth}
 {
 \caption[coeffl4]
 {\label{coeffl4}
\small
Coefficients in the polynomials $B(\mu)= \sum_{k=0}^{L^2} B_k(\mu) \, u^k$, 
for the four independent 2 by 2 block spin configurations on an 
$L=4$ lattice, majority rule.
}}
\end{center}
\end{table}

\begin{figure}
\begin{center}
\includegraphics[width=11cm]{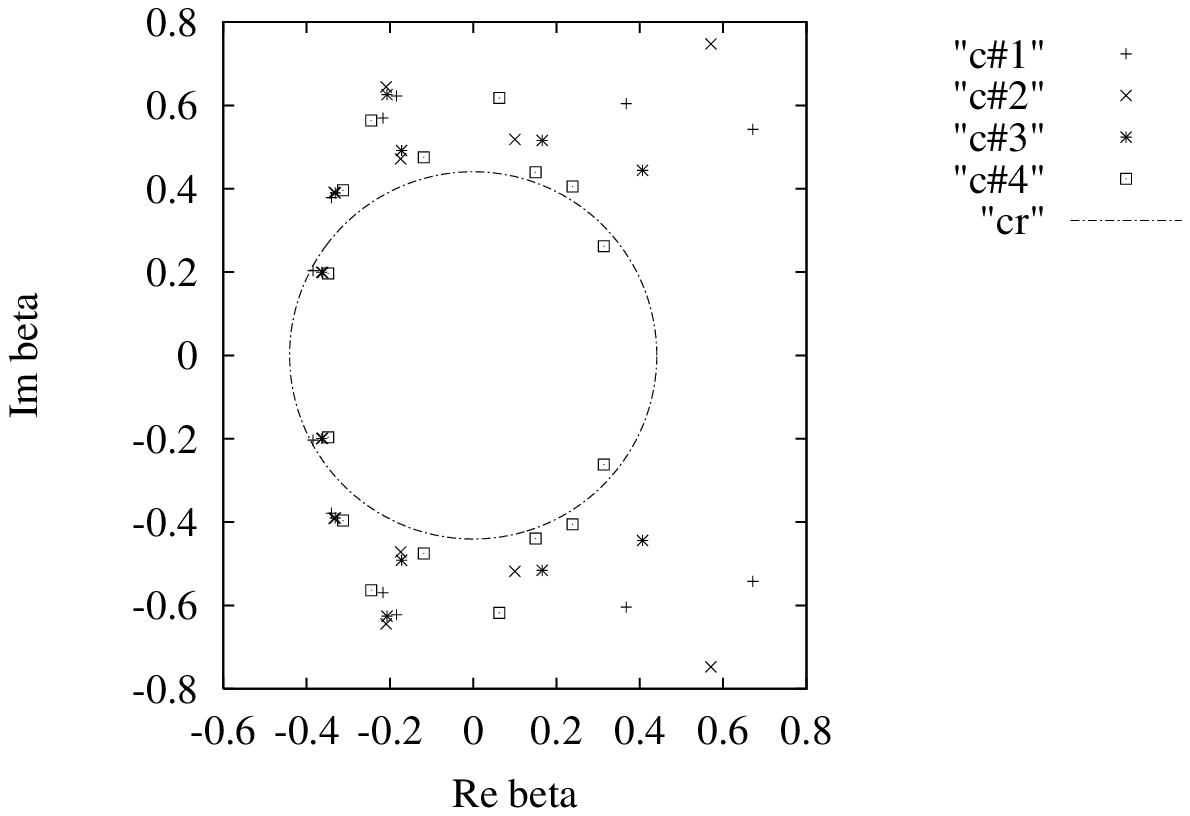}
\parbox[t]{.85\textwidth}
 {
 \caption[fig1]
 {\label{fig1}
\small
Zeroes of $B(\mu)$ for the four non-equivalent block spin configurations 
on a 2 by 2 block lattice, majority rule.
The circle has radius $\beta_c$.
}}
\end{center}
\end{figure}

The zeroes of these polynomials were determined with the help of the
computer algebra program MapleV. For zeroes $u_0$ not lying on the
negative real axis, we then computed the corresponding $\beta_0$
values through $\beta_0 = \frac14 \ln(u_0)$.  The distributions of
these numbers for $L=4$ are shown in figure~\ref{fig1}, with different
symbol code for the four block spin configurations. The figure also
contains a circle of radius $\beta_c$ around the origin.  Note that
the two zeroes closest to the critical point belong to configuration
$c\#4$.

\begin{figure}
\begin{center}
\includegraphics[width=11cm]{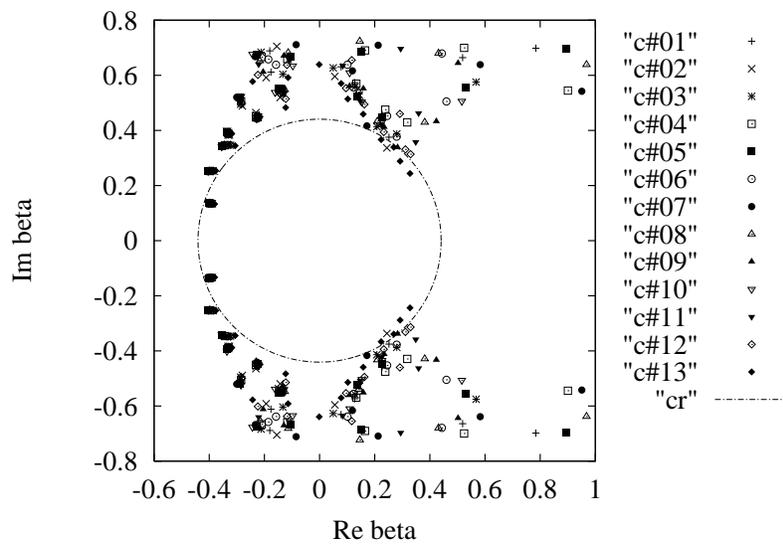}
\parbox[t]{.85\textwidth}
 {
 \caption[fig2]
 {\label{fig2}
\small
Zeroes of $B(\mu)$ for the 13 non-equivalent block spin configurations 
on a 3 by 3 block lattice, majority rule.
}}
\end{center}
\end{figure}

The results for the 3 by 3 block lattice are plotted in
figure~\ref{fig2}. The 13 non-equivalent block spin configurations are
specified in table~\ref{tabc6}.  One observes that again the zeroes
closest to the critical point belong to the fully magnetized
configuration ($c\#13$).

\begin{table}
\small 
\newcommand{\cffA}{$\begin{array}{|ccc|} 
\multicolumn{3}{c}{c\#01} \\
\hline 
+ & + & - \\[2mm] 
+ & + & - \\[2mm]
+ & - & - \\
\hline 
\end{array}$}
\newcommand{\cffB}{$\begin{array}{|ccc|} 
\multicolumn{3}{c}{c\#02} \\
\hline 
 + & + & - \\[2mm]       
 + & + & - \\[2mm]
 + & + & - \\
\hline          
\end{array}$}
\newcommand{\cffC}{$\begin{array}{|ccc|} 
\multicolumn{3}{c}{c\#03} \\
\hline 
 + & + & + \\[2mm]      
 + & - & - \\[2mm]
 + & - & - \\
\hline 
\end{array}$}
\newcommand{\cffD}{$\begin{array}{|ccc|} 
\multicolumn{3}{c}{c\#04} \\
\hline 
 + & - & + \\[2mm]     
 + & + & - \\[2mm]
 + & - & - \\
\hline 
\end{array}$}
\newcommand{\cffE}{$\begin{array}{|ccc|} 
\multicolumn{3}{c}{c\#05} \\
\hline 
 - & + & + \\[2mm]      
 + & + & - \\[2mm]
 + & - & - \\
\hline 
\end{array}$}
\newcommand{\cffF}{$\begin{array}{|ccc|} 
\multicolumn{3}{c}{c\#06} \\
\hline 
 + & + & + \\[2mm]       
 + & + & - \\[2mm]
 + & - & - \\
\hline 
\end{array}$}
\newcommand{\cffG}{$\begin{array}{|ccc|} 
\multicolumn{3}{c}{c\#07} \\
\hline 
 - & - & + \\[2mm]      
 + & + & - \\[2mm]
 + & + & - \\
\hline 
\end{array}$}
\newcommand{\cffH}{$\begin{array}{|ccc|} 
\multicolumn{3}{c}{c\#08} \\
\hline 
 + & - & + \\[2mm]      
 + & + & - \\[2mm]
 + & + & - \\
\hline 
\end{array}$}
\newcommand{\cffI}{$\begin{array}{|ccc|} 
\multicolumn{3}{c}{c\#09} \\
\hline 
 + & + & + \\[2mm]       
 + & + & - \\[2mm]
 + & + & - \\
\hline 
\end{array}$}
\newcommand{\cffJ}{$\begin{array}{|ccc|} 
\multicolumn{3}{c}{c\#10} \\
\hline 
 - & + & + \\[2mm]     
 + & - & + \\[2mm]
 + & + & - \\
\hline     
\end{array}$}
\newcommand{\cffK}{$\begin{array}{|ccc|} 
\multicolumn{3}{c}{c\#11} \\
\hline 
 + & + & + \\[2mm]     
 + & - & + \\[2mm]
 + & + & - \\
\hline 
\end{array}$}
\newcommand{\cffL}{$\begin{array}{|ccc|} 
\multicolumn{3}{c}{c\#12} \\
\hline 
 + & + & + \\[2mm]     
 + & + & + \\[2mm]
 + & + & - \\
\hline 
\end{array}$}
\newcommand{\cffM}{$\begin{array}{|ccc|} 
\multicolumn{3}{c}{c\#13} \\
\hline 
 + & + & + \\[2mm]     
 + & + & + \\[2mm]
 + & + & + \\         
\hline 
\end{array}$}
\small
\begin{center}
\begin{tabular}{cccc}
\cffA & \cffB  & \cffC & \cffD  \\[10mm]
\cffE & \cffF  & \cffG & \cffH  \\[10mm]
\cffI & \cffJ  & \cffK & \cffL  \\[10mm]
\cffM &        &       &        \\
 \end{tabular}
\parbox[t]{.85\textwidth}
 {
 \caption[tabc6]
 {\label{tabc6}
\small
The 13 non-equivalent block spin configurations on a 3 by 3 lattice. 
}}
\end{center}
\end{table}

\begin{figure}
\begin{center}
\includegraphics[width=11cm]{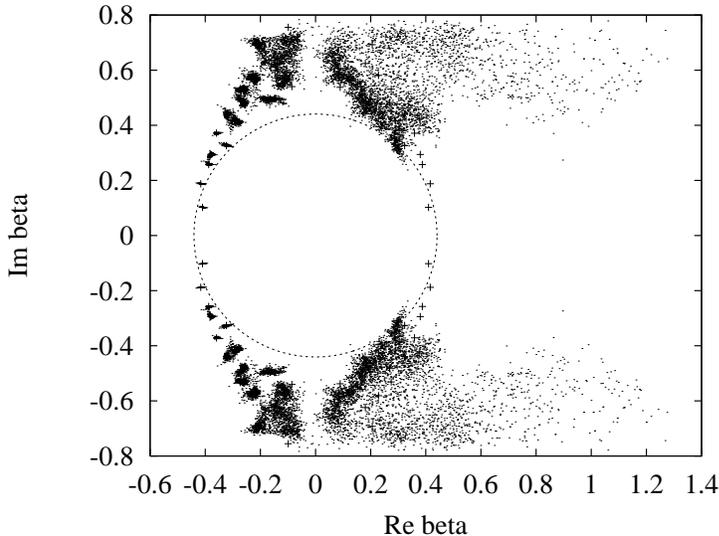}
\parbox[t]{.85\textwidth}
 {
 \caption[fig3]
 {\label{fig3}
\small
Zeroes of $B(\mu)$ for the 479 non-equivalent block spin configurations 
on a 4 by 4 block lattice, majority rule.
In addition, the zeroes of the full partition function on an 8 by 8 
lattice are shown (crosses).
}}
\end{center}
\end{figure}

\begin{figure}
\begin{center}
\includegraphics[width=11cm]{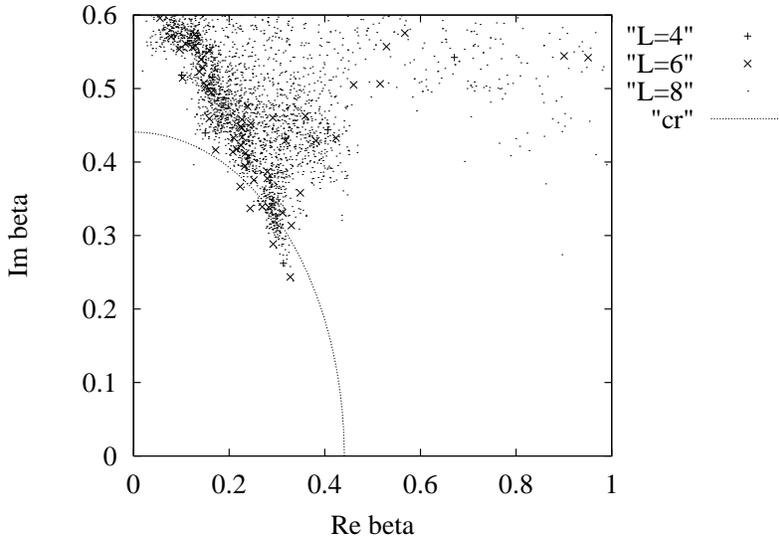}
\parbox[t]{.85\textwidth}
 {
 \caption[fig4]
 {\label{fig4}
\small
Common plot of the Boltzmann zeroes for $L=4$, 6, and 8, majority rule.
}}
\end{center}
\end{figure}

The zeroes of the 479 effective Boltzmannians on the 4 by 4 block
lattice are shown in figure~\ref{fig3}. The plot also shows (with
crosses) the zeroes of the full partition function.  It seems that the
Boltzmannian zeroes do not approach the real axis in the critical
region, whereas the full partition function zeroes do. To check this
in more detail, we compare the distribution of zeroes in the
critical region for the three available lattice sizes together, see
figure~\ref{fig4}.  The plot clearly demonstrates that the zeroes do
not move towards the real axis in the critical region.  One might 
conclude from this plot that there should exist in the
$L\rightarrow \infty$ limit a strip around the real $\beta$ axis
ranging from $\beta=0$ at least up to $\beta_c$ where the effective
Boltzmannian is free of $\beta$ zeroes. In this region it should thus
be possible to take the logarithm without danger. Furthermore, high
temperature (small $\beta$) expansions for the renormalization group
could by analytical continuation be used in the critical region.

There is another observation when comparing figures \ref{fig2} and
\ref{fig3}. With increasing $L$, more and more zeroes populate the
part of the plane with larger real part of $\beta$.  They are not
obviously approaching the real axis there, but we also cannot exclude
such a scenario.  Note that in the analysis of van Enter et
al.~\cite{vanEnter} the case of the 2D Ising majority rule was not
treated. It is therefore presently not clear whether in this case a
large $\beta$ pathology exists. We shall see in the next section 
that in case of the Kadanoff rule (where pathologies do exist) 
spurious zeroes seem to approach the axis at large positive 
$\beta$. 

\section{Results for Kadanoff Rule}

It has been observed many times that passing from a 
$\delta$-function block rule to a ``Gaussian smeared'' 
rule improves the locality and analyticity
properties of the effective Hamiltonian, see e.g.\ \cite{Hasenfratz}. 
In this section some results will be presented on the distribution 
of zeroes in the case of the Kadanoff rule eq.~(\ref{kadanoff}) for
$\omega=1$ and $\omega=2$. For finite $\omega$ there is a finite 
probability  that the block spin does not have the same sign 
as the majority of spins in the block: 
\begin{center}
\begin{tabular}{ccc}
$\sum_{i \in I} \sigma_I$  
& prob($\mu=1)_{\omega=1}$ 
& prob($\mu=1)_{\omega=2}$ \\   
\hline 
4 &  0.99966465 &  0.99999989  \\
2 &  0.98201379 &  0.99966465  \\         
\end{tabular}
\end{center}

\begin{figure}
\begin{center}
\includegraphics[width=11cm]{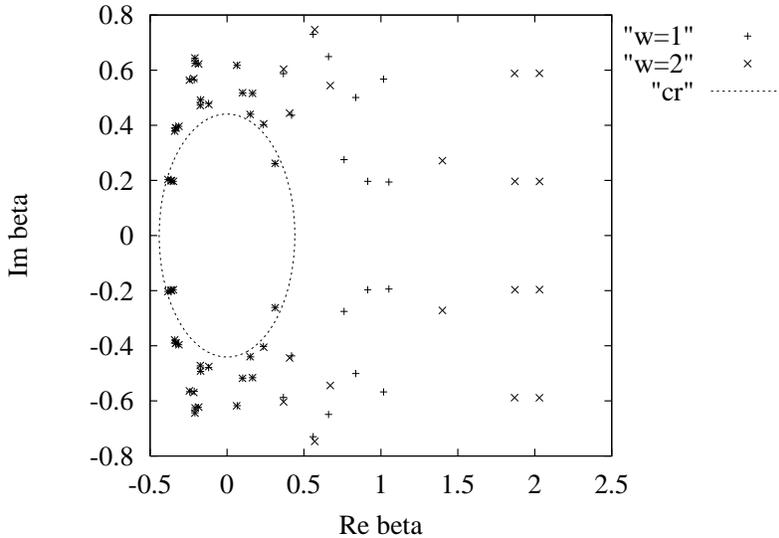}
\parbox[t]{.85\textwidth}
 {
 \caption[fig5]
 {\label{fig5}
\small
Zeroes of $B(\mu)$ for $L=4$, 
for the majority rule ($\omega=\infty$), and 
for the Kadanoff rules with $\omega=1$ and $\omega=2$.
}}
\end{center}
\end{figure}

\begin{figure}
\begin{center}
\includegraphics[width=11cm]{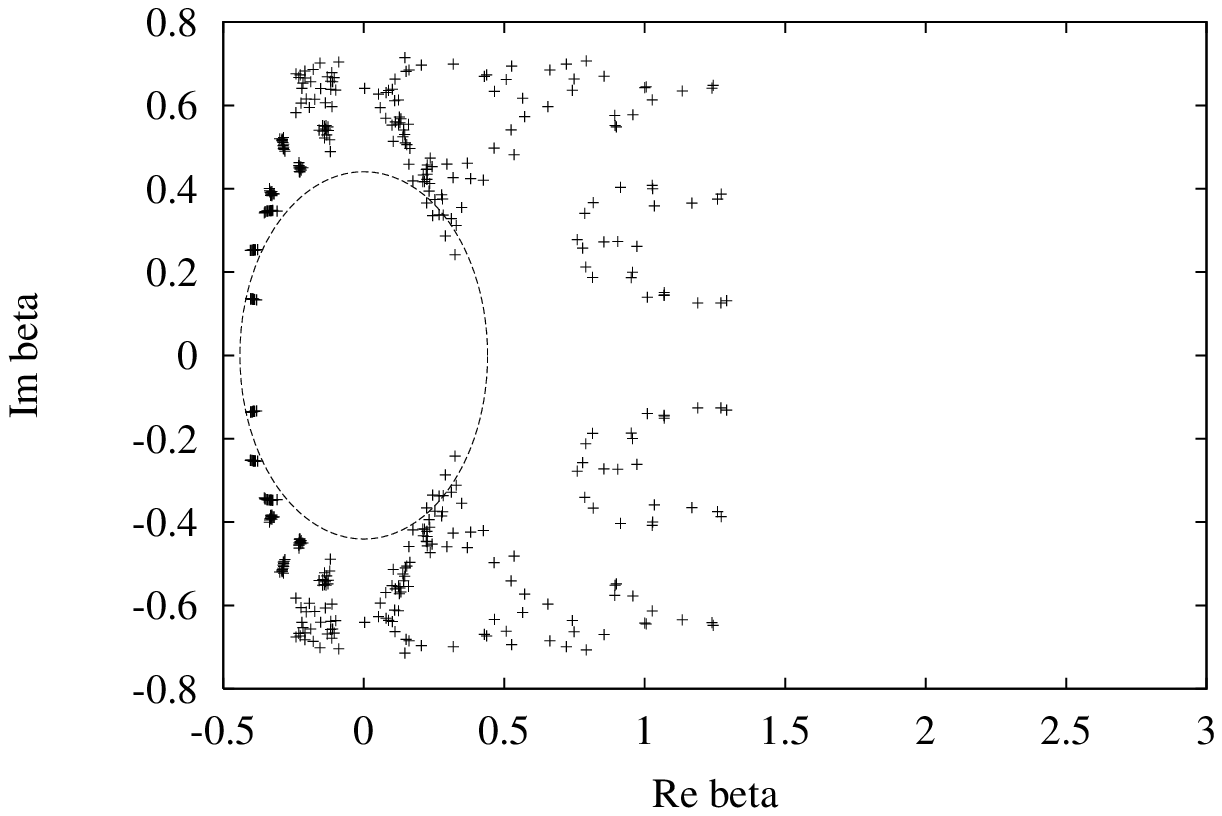}

\includegraphics[width=11cm]{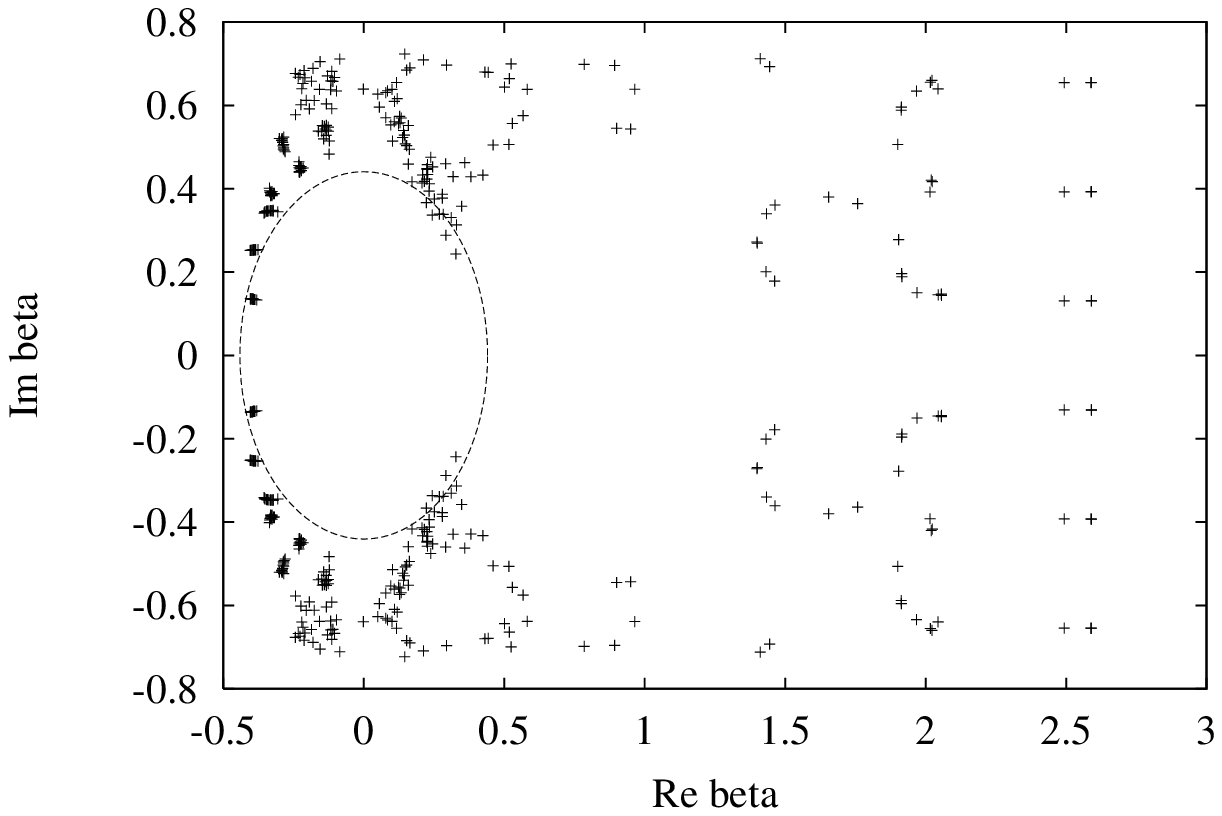}
\parbox[t]{.85\textwidth}
 {
 \caption[fig6]
 {\label{fig6}
\small
Zeroes of $B(\mu)$ for $L=6$, with Kadanoff rule. Top: $\omega=1$, 
bottom: $\omega=2$. 
}}
\end{center}
\end{figure}

In figures \ref{fig5} and \ref{fig6} 
we show the zeroes for $L=4$ and $L=6$, both for $\omega= 1$ 
and $\omega=2$. 
Obviously, most of the zeroes in the neighbourhood of the 
critical circle do not move very much. In fact they are already
very close to their majority rule values. 
However, compared with the $\omega=\infty$ case, extra zeroes 
appear that populate the region of larger real part of $\beta$. 
It might be interesting to note that the zeroes with the 
largest real part of $\beta$ belong to block spin configuration 
$c\#2$, followed by those of $c\#1$.
Another observation, also clearly seen in figure \ref{fig6},
is that the spurious zeroes move to the right when $\omega$
is increased. Most likely they are shifted to infinity when 
passing to the majority rule. 

In figure 7 the zeroes of all three lattice sizes, $L=4$, 6, and 8 are
plotted for the case $\omega=1$.\footnote{For $L=8$, some 16 of the
479 block spin configurations were not taken into account because
the search for the zeroes of the corresponding polynomials suffered
from numerical instabilities.}  A careful inspection reveals that
the zeroes in the right half plane {\em do} move towards the real axis
when the lattice size is increased. They might thus very well be
reflecting the large $\beta$ pathologies.

\begin{figure}
\begin{center}
\includegraphics[width=11cm]{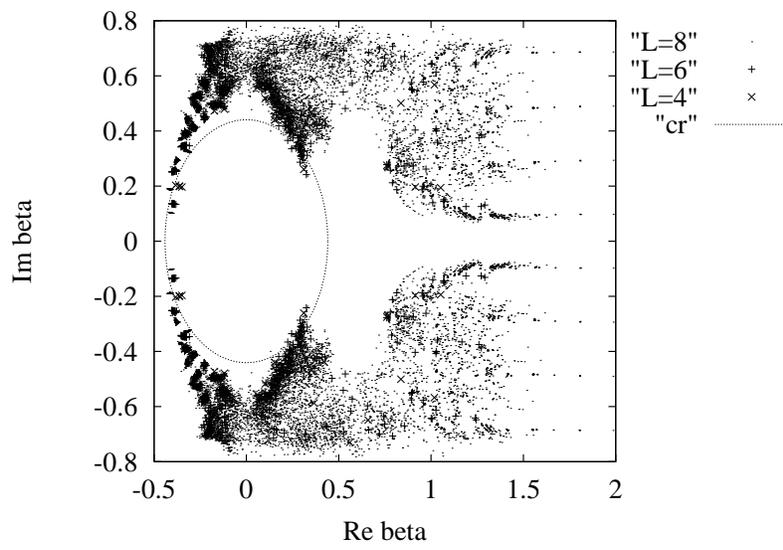}
\parbox[t]{.85\textwidth}
 {
 \caption[fig7]
 {\label{fig7}
\small
Zeroes of $B(\mu)$ for $L=4$, 6, and 8, with Kadanoff rule, $\omega=1$.
}}
\end{center}
\end{figure}

\section*{Conclusions}

The distribution of zeroes of effective Boltzmannians was studied for
2D Ising systems. Both in the
case of the majority and the Kadanoff rule a finite region around the
critical point stays free of zeroes, also when the volume is
increased. 
In case of the Kadanoff rule, however, zeroes populate the
right half plane and approach the real axis at large $\beta$.
They might be
related to the pathologies of a number of Ising renormalization groups
discussed in the literature.  It would be very interesting to
understand this relation (if it exists). Furthermore, one should try
to understand the origin of the extra zeroes that appear for finite
$\omega$.  Van Enter et al.\ were not able to extend their analysis of
pathologies to $\omega=\infty$ in two dimensions.  The present study
shows that the distribution of
large $\beta$ zeroes changes significantly when $\omega$ becomes finite.
The present findings do not, however, exclude the possibility
of large $\beta$ pathologies of the 2D Ising majority rule
renormalization group.

\end{document}